# A Pre-Geometric Model Exhibiting Physical Law

**Charles Francis**

**Abstract:**
   We describe a substructure for matter in which metric relations are not assumed as properties of a physical manifold, and instead the metric of general relativity is found in configurations of particle interactions. The 'state' of a particle is the set of possible configurations of surrounding particles. States are thus non-local, while particles are local. Kets are not states, or descriptions of states, but merely names for states. Ket space is a naming system constructed by an individual observer from the available information in such a way that the principle of superposition and the probability interpretation are definitional truisms, not physical laws. In this paper the treatment is developed as far as showing that the solution to the discrete equations describing the structure should be embedded into covariant continuous equations, and we remark that this leads to a requirement for four dimensions and spin in the organisation of a structure consisting solely of particle interactions, although neither dimensionality nor spin are assumed properties of particles.

Charles Francis
Clef Digital Systems Ltd
Lluest, Neuaddlwyd
Lampeter
Ceredigion
SA48 7RG

24/9/99

# A Pre-Geometric Model Exhibiting Physical Law

## 1   The Fabric of Space-time

There have been previous suggestions that at a fundamental level of physical law, variables such as time should be discrete [1], and there are many references in the literature on the potential quantisation of gravity which suggest that this may be so [2]. Just as we recognise that the surface of a sheet of paper is fibrous and non-geometrical, we may also conceive that ontological space is not smooth or continuous. The microscopic structure of a manifold such as that used in general relativity is described by the theory of limits, which are described in mathematics but have no direct empirical basis except as approximation to finite procedures, and should be seen in the light of Riemann's mathematical definition [3] which deliberately ignores the question of whether the manifold represents something ontological. Riemann had this to say (Clifford's translation, quoted in [4])

Either, therefore, the reality which underlies space must form a discrete manifold, or we must seek the ground of its metric relations outside it.

Newton himself was aware of an inherent circularity in the description of mechanics in terms of absolute space and time, since he wrote, on page 1 of his preface to the first edition of the Principia,

The description of right lines and circles, upon which geometry is founded, belongs to mechanics. Geometry does not teach us to draw these lines, but requires them to be drawn.

The suggestion is that absolute time and space were intended to be mathematical concepts, not physical entities, and that his mathematical principles of physics can be expected to hold only to the extent that the values they describe can be approximated in mechanics. This answer is again indicated by the orthodox interpretation of quantum mechanics, [5][6], which can be understood as suggesting that there is no real manifold. According to Dirac [7]

The expression that an observable 'has a particular value' for a particular state is permissible in quantum mechanics in the special case when a measurement of the observable is certain to lead to the particular value, so that the state is an eigenstate of the observable...In the general case we cannot speak of an observable having a value for a particular state, but we can .... speak of the probability of its having a specified value for the state, meaning the probability of this specified value being obtained when one makes a measurement of the observable.

Thus we cannot say that the property of position necessarily exists between measurements, and we cannot justify the assumption of a physical manifold, or pre-existent continuum modelled by $\mathbb{R}^n$ into which matter can be placed. This paper will examine an interpretation of physical law which provides the ground for metric relations which are not a reflection of an ontological manifold underlying space, but rather of relationships found in particle interactions. The motivation for this is that quantum electrodynamics has shown that the exchange of photons is responsible for the electromagnetic force, and so for all the structures of matter in our macroscopic environment. But as Bondi pointed out, the exchange of photons is also the process used to measure space-time co-ordinate systems by the radar method [8]. It is not unreasonable, therefore, to postulate that photon exchange generates all the geometrical relationships in the macroscopic environment, just as it generates these relationships in the results of measurement by radar.

The purpose of the present paper is to examine the implications of such an assumption and develops the model as far as the requirement for dimensionality and the Dirac equation. The model has been further developed to a discrete reformulation of quantum electrodynamics [9] and Maxwell's equations, and differs in its predictions from the standard model in so far as Feynman rules give finite and unambiguous results without renormalisation.

## 2 Particles

We seek to investigate the properties of a structure consisting solely of abstract particles and in which all relationships found within the structure are derived from internal properties of the structure. Later abstract particles will be identified with the elementary particles of physics, but they are defined by the properties 2.1 to 2.4. Where there is no ambiguity we will refer to abstract particles simply as particles.

2.1 *Particles are indivisible*

This paper will investigate only two types of particles, electrons and photons, though the treatment is general enough to describe other particles. We will find that these abstract electrons and photons have properties very like those of physical electrons and photons.

If particles are to model physics they must have behaviour with respect to time. Since the structure is to consist only of particles, we assume that time is a property of the particles themseves. The fundamental unit of time for a particle is called the chronon, after its name in antiquity. There may be a different chronon for each elementary particle. We assume that the chronon is very much smaller than the unit given by any practical clock, and that for practical purposes conventional measures of time can be regarded as (large) whole numbers of chronons.

**Definition:** Each segment of a time line is an ordered set homomorphic to a segment of $\mathbb{N}$.

2.2 *Each particle has a time line*

Intervals in the time line of a particle are not assumed to be directly observable. A macroscopic clock is determined by many particles, so it will be assumed that

*2.3 Macroscopic time is a statistical composition of the time lines of particles.*

**Definition:** Let $\chi \in \mathbb{N}$ be the scaling factor to chronons from conventional units of time.

To derive some behaviour for the model we assume interactions between photons and electrons. This is assumed to take place in the simplest possible way.

2.4 *Photons can be emitted/absorbed by electrons*

The distinction between emission/absorption is not available to a single interaction in the absence of context; it requires a comparison between the time line of one particle and the time line of another and is only possible in the context of several interactions.

Since visualisation involves the awareness of geometry, it is clear that, strictly, the pre-geometric properties of matter cannot be visualised. Nonetheless, interactions between particles can be illustrated diagrammatically. Electrons are shown as dashed lines, the dashes representing chronons, the fundamental discrete unit of time (figure 2.5). Photons are shown by continuous grey lines. Interactions are shown as nodes joining the end of a photon line to a point of an electron line (figure 2.6).

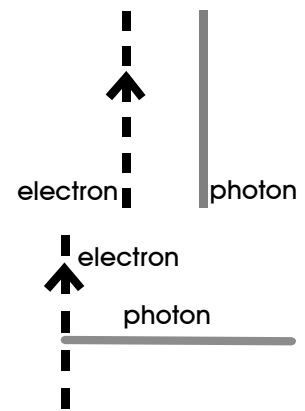

**Figure 2.5:** *A segment of the time line of an electron may be represented as an ordered set of dashes, each dash representing one fundamental unit (chronon) of time. The arrow denotes the ordering. The time line of a photon consists of only two points joined by a continuous line shown in grey.*

**Figure 2.6:** *An interaction in which a photon is emitted/absorbed by an electron.*

We can imagine a system of such particles in the form of a diagram provided that it is understood that the space between the lines of the diagram has no practical meaning. The geometrical properties of space-time depend on internal relationships between dashes and nodes in the diagram, not on the geometrical properties of the drawing. Thus, in figure 2.7, the increasing lapsed time for the return of a photon indicates that the particles are moving away from each other.

**Figure 2.7:** *A fabric of particle interactions.*

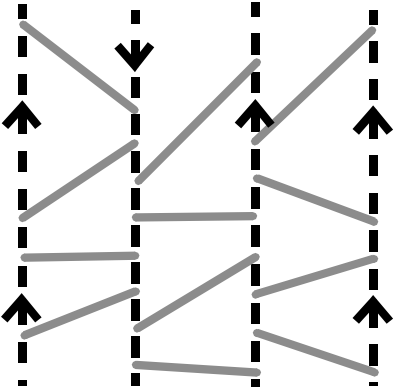

**Definition:** When a photon is absorbed and in the next chronon one is emitted, we will say that a photon is reflected, and we will describe the reflection as an event.

We can now define relationships between time-lines and events whenever a photon passes from one electron to another and is reflected back to the first electron, as shown in figure 2.8.

**Definition:** The distance of an event is half the lapsed time (in chronons) for a photon to go from the clock to the event and return to the clock. The time at which the photon is reflected is the mean time between when it is sent and when it returns.

**Figure 2.8:** *A "stitch in space-time" defines a relationship between events and time-lines whenever a photon passes between two electrons, and one immediately returns.*

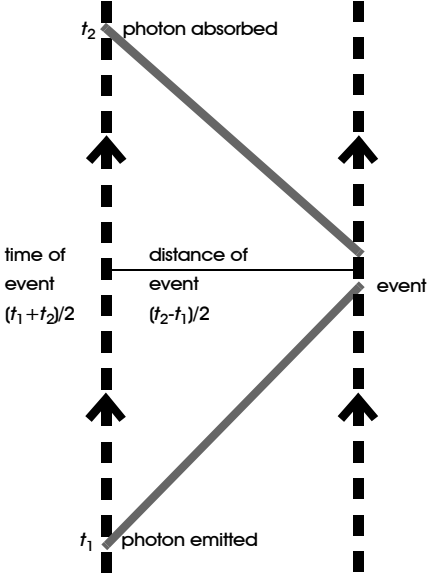

Although figure 2.8 closely resembles the radar method of measuring space-time co-ordinates, it is not radar, but describes a "hidden" subatomic process in which the values $t_1$ and $t_2$ are inherently unknown. The radar method (figure 2.9) uses macroscopic time, but otherwise the process of photon exchange is identical to the process in figure 2.8. Obviously we cannot empirically analyse individual exchanges of photons. But we can statistically analyse the effect of many such exchanges. Macroscopic behaviour will be predicted from the statistical analysis of the behaviour of many particles. In systems of many particles such as we generally observe, photons are constantly exchanged. Since the process of photon exchange is the same as we use in radar, the average behaviour of a system in which there are many such exchanges should obey the geometrical relationships found using radar to measure distance. Ignoring the number of space dimensions, it is straightforward to derive the formulae of relativity from the radar method [8][9]. Thus Minkowsky space-time can be construed as a composition of the primitive space-times associated with particles.

While these exchanges take place all the time in the subatomic structure of matter and geometrical relationships are found almost everywhere, we do not assume that it is always possible to find a metric based on photon exchange. Indeed, where there are few interactions a metric may not be defined at all. We also expect the metric to break down inside a black hole or at the big bang.

**Figure 2.9:** *Comparison with the radar method of determining co-ordinates shows the replacement of a particle's time line with time on a macroscopic clock.*

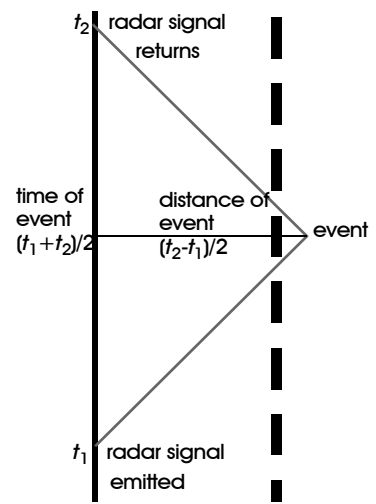

It will be observed that the model does not describe position as a property of a particle, but as a property pertaining to a configuration of particles in which the process shown in figure 2.9 (i.e. radar) has been used to generate a space-time co-ordinate, or in which some other process has been used which can be shown to generate the same result. It is clearly impossible to determine every detail of the configuration, since each detail would require an additional measurement, which would mean a larger configuration of particles with new unknown parts. Hence there is a residual level of uncertainty, which can never be removed by experiment. A macroscopic reference frame constitutes some form of average of the behaviour of individual particles and each individual particle is in part responsible for the generation of the macroscopic reference frame, so individual particles do not in general have exact position and uncertainty in position will be a feature of the description of elementary particles. We seek to establish that this inherent uncertainty is reflected in the laws of quantum mechanics, and thus that the uncertainty principle can be regarded as a consequence of an extension of the principle of relativity, namely that position, as well as motion, is relative.

It is clear that Minkowsky space-time will derive statistically from the microscopic properties of individual exchanges, but it is immediately possible to see an "inaccuracy" due to the time required for absorption and emission of photons. The metric cannot distinguish the times of absorption and emission of the reflecting photon, so there is a singularity at the reflection. It is not a true singularity, since, with a stastically derived metric, it is averaged out over particle interactions, and appears in macroscopic space-time as curvature not as a cusp. It is obvious that the curvature generated by the singularity is proportional to the chronon, which is the time componant of a vector and so proportional to mass-energy. Thus we see that, when it exists and ignoring spacial dimension for the time being, the metric is that described by Einstein's Field equation for general relativity, without the cosmological constant [10][11]. For simplicity of exposition we will refer to only one space dimension, but the treatment is general enough to apply to $n$-dimensions with curvature.

## 3    Ket Space

The determination of properties predicted on the assumption that the universe consists solely of particles depends on the analysis of measurement. Measurement is not taken to imply measurement *of* a pre-existent quantity, but is simply the generation of a value from the interactions of a system consisting of both an apparatus and a particle (or subsystem) under study. Although the actual configuration of particles in a structure such as that shown in figure 2.7 is assumed to be objective as demanded by realism, it is unknown to all observers, and we cannot frame observable physical law directly in terms of particle configurations. The definition of a state which can be described in empirical law depends on the information subjectively available to an observer about the objective configuration of particles, and whatever information is available allows many possible configurations of particles.

**Definition:** The state of a particle is the set of possible configurations of surrounding particles given the information available from observation.

The measurement of time and position is sufficient for the study of many other physical quantities; a classical measurement of velocity may be reduced to a time trial over a measured distance, and a typical measurement of momentum of a particle involves plotting its path in a bubble chamber. In chronons, the result of a measurement of position at a particular time is a point in $\mathbb{N}$. In practice there is also a bound on the magnitude of the result, so we may take the results of measurement of position to be in a finite region $N = (-\nu, \nu] \subset \mathbb{N}$ for some $\nu \in \mathbb{N}$, where $(-\nu, \nu] = \{x \in \mathbb{N} | -\nu < x \leq \nu\}$ $\nu$ is not a bound on the size of the model and merely has to be large enough to be able to say with certainty that N contains any particle under study, i.e. the position function of the particle vanishes outside of N. Although only one dimension is being considered at this point, we denote position with bold type for consistency of notation when more dimensions are introduced, and we regared $\boldsymbol{x}$ as vector.

**Definition:** $\forall \boldsymbol{x} \in N$, $|\boldsymbol{x}\rangle$ is a label corresponding to a measurement of position of a single particle with result $\boldsymbol{x}$. $|\boldsymbol{x}\rangle$ is called a position ket.

In the absence of information, we cannot describe the actual configuration of particles; kets are names or labels for states, not descriptions of matter. The significance is that the principle of superposition will be introduced as a definitional truism in a naming system, not as a physical assumption. Although kets are not states, but merely names for states, we loosely refer to kets as states in keeping with common practice when no ambiguity arises. It is worth remarking that while particles have no definable size, 'states' are non-local in space and time, consisting of configurations of many particles.

In a typical measurement in quantum mechanics we study a particle in near isolation. The implication is that there are too few ontological relationships to generate the property of position. Then position does not exist prior to the measurement, and the measurement itself is responsible for introducing interactions which generate position. In this case, prior to the measurement, the state of the system is not labeled by a position ket. We seek a means of categorising such states. In scientific measurement we set up many repetitions of the system, and record the frequency of each result. Probability is simply a prediction of frequency, so a mathematical model must associate a probability for each possible result. The probability of a given result can be used to attach a label to the state of the system in the following way:

**Definition:** Let $\mathbb{H}_0$ be the set of kets $\mathbb{H}_0 = \{|\boldsymbol{x}\rangle | \boldsymbol{x} \in N\}$

**Definition:** Construct a vector space, $\mathbb{H}$, over $\mathbb{C}$, with basis $\mathbb{H}_0$. The members of $\mathbb{H}$ are called kets.

This is trivial because $\mathbb{H}_0$ is finite. $\mathbb{H}$ has dimension $2\nu$. $\mathbb{H}$ is isomorphic to as the set of $2\nu \times 1$ matrices generated by the operations of addition and multiplication by a scalar from basis kets represented by a matrix containing a 1 and with all other entries equal to 0. We can use the infinite field of complex numbers because this is purely a mathematical device, like $\sqrt{-1}$, requiring no physical justification.

**Definition:** $\forall |f\rangle, |g\rangle \in \mathbb{H}$, the braket $\langle g|f\rangle$ is the hermitian form on $\mathbb{H}$ defined by its action $\mathbb{H}_0$

3.1 $\qquad \forall x, y \in \mathrm{N}, \langle x|y\rangle = \delta_{xy}$

**Definition:** The position function $f(x)$ of the ket $|f\rangle \in \mathbb{H}$ is the function $f: \mathrm{N} \to \mathbb{C}$ defined by

3.2 $\qquad \forall x \in \mathrm{N}, x \to f(x) = \langle x|f\rangle$

Later the position function will be identified with the restriction of the wave function to $\mathbb{N}$, but we use the term position function, because it is discrete, and because a wave is not assumed. We observe that

3.3 $\qquad P(x) = \dfrac{|\langle f|x\rangle|^2}{\langle f|f\rangle}$

has the properties of a probability distribution, and we associate the ket $|f\rangle$ with any state such that $\forall x \in \mathrm{N}$ the probability that a measurement of position has result $x$ is given by $P(x)$. In this association kets are not states, or descriptions of states, but merely labels for states. The same ket may label different states provided only that 3.3 is satisfied. It is routine to prove that

3.4 $\qquad \forall |f\rangle \in \mathbb{H}, |f\rangle = \sum_{x \in \mathrm{N}} |x\rangle\langle x|f\rangle$

and that the braket is given by

3.5 $\qquad \langle g|f\rangle = \sum_{x \in \mathrm{N}} \langle g|x\rangle\langle x|f\rangle$

3.4 is true for all $|f\rangle$, and hence we can define an operator expression known as the resolution of unity

3.6 $\qquad \sum_{x \in \mathrm{N}} |x\rangle\langle x| = 1$

Vector space extends the naming system from $\mathbb{H}_0$ to $\mathbb{H}$ and introduces intuitive logical operations between uncertain propositions. Indeed the scalar product (and in particular the position function) can be understood as a truth value in a many valued logic [13][9]. Addition corresponds to logical OR, and multiplication by a scalar gives an intuitive idea of weighting due to the level of certainty in each option given to logical OR. Multiplication by a scalar only has meaning as a weighting between alternatives and we have $\forall |f\rangle \in \mathbb{H}, \forall \lambda \in \mathbb{C}$ such that $\lambda \neq 0$, $\lambda|f\rangle$ labels the same state (or set of states) as $|f\rangle$.

## 4 Momentum Space

**Definition:** Momentum space is $\mathrm{M} = (-\pi, \pi] = \{p \in \mathbb{R} | -\pi < p \leq \pi\}$. The elements of momentum space are called momenta.

**Definition:** For each value of momentum $p \in \mathrm{M}$, define a ket $|p\rangle$, known as a plane wave state, by the position function

4.1 $\qquad \langle x|p\rangle = \left(\dfrac{1}{2\pi}\right)^{\frac{1}{2}} e^{-ix \cdot p}$

It will be found that $p$ gives rise to classical momentum [9]. Clearly the cardinality of the plane wave states is greater than the cardinality of $\mathbb{H}_0$ so plane waves are not a basis. The dot product is used in 4.1 so that the treatment is general enough to incorporate curvature in $n$ dimensions as discussed in section 2, *Particles* (if $g$ is the metric tensor, restricted to the synchronous surface defined by radar, we will have $x \cdot p = x_i g^{ij} p_j$, where I have used the standard convention on raising and lowering indices).

The expansion of $|p\rangle$ in the basis $\mathbb{H}_0$ is calculated by using the resolution of unity, 3.6

$$4.2 \qquad |p\rangle = \sum_{x \in \mathbb{N}} |x\rangle\langle x|p\rangle = \left(\frac{1}{2\pi}\right)^{\frac{1}{2}} \sum_{x \in \mathbb{N}} e^{-ix \cdot p}|x\rangle$$

**Definition:** $\forall(|f\rangle \in \mathbb{H})$ define the transform, $F:\mathrm{M} \to \mathbb{C}$, also called the momentum space function

$$4.3 \qquad F(p) = \langle p|f\rangle$$

**Lemma:** $F$ can be expanded as a trigonometric polynomial

$$4.4 \qquad F(p) = \sum_{x \in \mathbb{N}} \langle p|x\rangle\langle x|f\rangle = \left(\frac{1}{2\pi}\right)^{\frac{1}{2}} \sum_{x \in \mathbb{N}} \langle x|f\rangle e^{ix \cdot p}$$

**Proof:** Apply 3.6, 4.1 and 4.3

**Lemma:**

$$4.5 \qquad \forall x, y \in \mathbb{N}, \quad \int_{-\pi}^{\pi} dp\, e^{-i(x-y) \cdot p} = \begin{cases} 2\pi & \text{if } y = x \\ 0 & \text{otherwise} \end{cases}$$

**Proof:** Straightforward trigonometry.

The position function can be found from the Fourier coefficient

$$\left(\frac{1}{2\pi}\right)^{\frac{1}{2}} \int_{\mathrm{M}} dp\, F(p) e^{-ix \cdot p} = \frac{1}{2\pi} \int_{\mathrm{M}} dp \sum_{y \in \mathbb{N}} \langle x|f\rangle e^{iy \cdot p} e^{-ix \cdot p} \qquad \text{by 4.4 and 4.1}$$

$$4.6 \qquad\qquad\qquad = \langle x|f\rangle \qquad\qquad\qquad \text{by 4.5}$$

Rewriting 4.6 in the notations of 4.3 and 4.1

$$4.7 \qquad \langle x|f\rangle = \int_{\mathrm{M}} dp\, \langle x|p\rangle\langle p|f\rangle$$

Then $\forall |f\rangle, |g\rangle \in \mathbb{H}$

$$\langle g|f\rangle = \sum_{x \in \mathbb{N}} \langle g|x\rangle\langle x|f\rangle \qquad \text{by 3.6}$$

$$= \int_{\mathrm{M}} dp \sum_{x \in \mathbb{N}} \langle g|x\rangle\langle x|p\rangle\langle p|f\rangle \qquad \text{by 4.7}$$

$$4.8 \qquad\qquad = \int_{\mathrm{M}} dp\, \langle g|p\rangle\langle p|f\rangle \qquad \text{by 3.6}$$

4.8 is true for all $|f\rangle$ and $|g\rangle$, and hence we can define a second operator expression known as the resolution of unity

$$4.9 \qquad \int_{\mathrm{M}} dp\, |p\rangle\langle p| = 1$$

It follows immediately that

$$4.10 \qquad \langle q|f\rangle = \int_{\mathrm{M}} dp\, \langle q|p\rangle\langle p|f\rangle$$

So $\langle p|q\rangle$ has the effect of a Dirac delta function on the test space of momentum space functions.

**Definition:** The delta function is

$$4.11 \qquad \delta:\mathrm{M} \to \mathbb{C} \qquad \delta(p - q) = \langle q|p\rangle$$

Explicitly, calculating $\langle p | q \rangle$ directly from 4.2

$$4.12 \qquad \delta(\boldsymbol{p} - \boldsymbol{q}) = \frac{1}{2\pi} \sum_{x \in N} e^{-i\boldsymbol{x} \cdot (\boldsymbol{p} - \boldsymbol{q})}$$

Clearly, for finite N, is a well defined; its property as a delta function is given in 4.10.

The lack of symmetry between momentum space and co-ordinate space reflects the fact that position is closely associated with a fundamental structure (figure 2.8), whereas momentum is a construction. The dependency of momentum space functions on N is irrelevant since it only effects kets with position functions exhibiting a sharp cutoff at the boundary of N. These are not considered here, and it is always possible to exclude them by increasing the value of ν. N is bounded so it is not possible to define unlimited space translation, but N is large enough to contain any particle under study, and can be taken larger without loss of generality. Under a space translation, $z$, of the co-ordinate system such that the particles under consideration certainly remain in N, 4.6 becomes

$$4.13 \qquad \forall \boldsymbol{x} \in N \quad \langle \boldsymbol{x} - \boldsymbol{z} | f \rangle = \begin{cases} \frac{1}{2\pi} \int_M d\boldsymbol{p}\, F(\boldsymbol{p}) e^{i\boldsymbol{z} \cdot \boldsymbol{p}} e^{-i\boldsymbol{x} \cdot \boldsymbol{p}} & \text{if } \boldsymbol{x} - \boldsymbol{z} \in N \\ 0 & \text{otherwise} \end{cases}$$

By 4.8, multiplication of the momentum space functions by $e^{i\boldsymbol{p} \cdot \boldsymbol{z}}$ is a homomorphic correspondence, and by 4.13 it is equivalent to space translation, $z$, of the co-ordinate system in the subspace of kets for states of particles which are certainly in N both before and after the translation.

## 5   Multiparticle States

**Definition:** The vector space, $\mathbb{H}^n$, of kets for naming multiparticle states of particles of the same type is defined by

*i.*   $\mathbb{H}^0 = \{\lambda | \rangle : \lambda \in \mathbb{C}\}$ i.e. the space containing only the empty ket, a name for a state of no particles, (the vacuum state). It is trivial that $\mathbb{H}^0$ is a one dimensional vector space isomorphic to $\mathbb{C}$, so we can identify $\mathbb{H}^0 = \mathbb{C}$. The empty ket is normalised to

$$5.1 \qquad \langle | \rangle = 1$$

*ii*   $\mathbb{H}^1 = \mathbb{H}^0 \oplus \mathbb{H}$ Clearly a one particle state cannot be a no particle state, so by the association of the braket with probability (3.3)

$$5.2 \qquad \forall |f\rangle \in \mathbb{H} \quad \langle | f \rangle = 0$$

*iii.*   For $n \in \mathbb{N}, n > 0$ $\mathbb{H}^n = \bigotimes_n \mathbb{H}^1$ (the external direct product).

Thus the elements of $\mathbb{H}^n$ are ordered *n*-tuples such that addition is given by

$$5.3 \qquad (|f_1\rangle, \ldots, |f_n\rangle) + (|g_1\rangle, \ldots, |g_n\rangle) = (|f_1\rangle + |g_1\rangle, \ldots, |f_n\rangle + |g_n\rangle)$$

and multiplication by a scalar $\lambda \in \mathbb{C}$ is given by

$$5.4 \qquad \lambda(|f_1\rangle, \ldots, |f_n\rangle) = (\lambda|f_1\rangle, \ldots, \lambda|f_n\rangle)$$

For the states $|f\rangle = (|f_1\rangle, \ldots, |f_n\rangle)$ and $|g\rangle = (|g_1\rangle, \ldots, |g_n\rangle)$ the braket is given by

$$5.5 \qquad \langle |f_1\rangle, \ldots, |f_n\rangle | |g_1\rangle, \ldots, |g_n\rangle \rangle = \prod_{i=1}^{n} \langle f_i | g_i \rangle$$

which is required for the probabilility interpretation, 3.3, if each of the particles is independent.

Hence, by 3.1, $\forall x^i \in N$, $i = 1, ..., n$ the basis $(|x^1\rangle, ..., |x^n\rangle)$ is normalised such that

5.6 $$\langle\langle|y^1\rangle, ..., |y^n\rangle\||x^1\rangle, ..., |x^n\rangle\rangle\rangle = \prod_{i=1}^{n} \delta_{y^i x^i}$$

**Definition:** Let $\mathbb{H}_0^n = \bigotimes_n \mathbb{H}_0 \cup \mathbb{H}^0$. Clearly $\mathbb{H}_0^n$ is a basis of $\mathbb{H}^n$.

**Definition:** The space of all particles of the same type is $\mathbb{H}^N$ where $N \in \mathbb{N}$ is larger than the number of particles in the universe.

**Corollary:** The statement that we can take a value of $N$ greater than any given value is the definition of an infinite sequence, so, in effect the space of all particles of the same type is $\mathbb{H}^\infty$

**Corollary:** $\forall i, n \in \mathbb{N}$, such that $0 < i < n$, $\mathbb{H}^i \subset \mathbb{H}^n$ is an isomorphic embedding under the mapping

5.7 $$\mathbb{H}^i \to \mathbb{H}^n : \quad (|f_1\rangle, ..., |f_i\rangle) \to (|f_1\rangle, ..., |f_i\rangle, |\rangle, ..., |\rangle)$$

**Definition:** The space of all particles is $\mathscr{H} = \bigotimes_\gamma \mathbb{H}_\gamma^\infty$ where $\gamma$ runs over every type of particle.

## 6 Creation Operators

The creation of a particle in an interaction is described by the action of a creation operator. Creation operators incorporate the idea that particles of the same type are identical, so that when a particle is created it is impossible to distinguish it from any existing particle of the same type. The definition removes arbitrary phase and normalises the two particle state to coincide with 5.6.

**Definition:** $\forall |x\rangle \in \mathbb{H}_0^1$ the creation operator $|x\rangle$ is defined by $\forall |y\rangle \in \mathbb{H}_0^1$

$$|x\rangle : |y\rangle \to |x\rangle|y\rangle = |x;y\rangle$$

6.1 $$= \frac{1}{\sqrt{2}}[(|x\rangle, |y\rangle) + \kappa(|y\rangle, |x\rangle)]$$

where $\kappa \in \mathbb{C}$ is to be determined.

**Definition:** The bra corresponding to $|x;y\rangle \in \mathbb{H}^2$ is designated by $\langle x;y| \in \mathbb{H}^2$

Now, by 5.5 and 6.1 $\forall x, y \in N$

$$\langle x;y|x;y\rangle = \tfrac{1}{2}[\langle x|x\rangle\langle y|y\rangle + \kappa^2\langle x|x\rangle\langle y|y\rangle + 2\kappa\langle x|y\rangle\langle y|x\rangle]$$

6.2 $$= \tfrac{1}{2}[(1 + \kappa^2) + 2\kappa\delta_{xy}^2]$$

The order in which particles are created can make no difference to the state, so

6.3 $\quad \exists \lambda \in \mathbb{C}$ such that $|x;y\rangle = \lambda|y;x\rangle$

Thus, by direct application of 5.5 and 6.1

$$\langle x;y|x;y\rangle = \lambda\langle x;y|y;x\rangle$$

$$= \tfrac{1}{2}\lambda[\kappa\langle x|x\rangle\langle y|y\rangle + \kappa\langle x|x\rangle\langle y|y\rangle + (1+\kappa^2)\langle x|y\rangle\langle y|x\rangle]$$

6.4 $$= \tfrac{1}{2}\lambda[2\kappa + (1+\kappa^2)\delta_{xy}^2]$$

Comparison of 6.2 with 6.4 gives

6.5 $\quad 1 + \kappa^2 = 2\lambda\kappa \quad$ and $\quad \lambda(1 + \kappa^2) = 2\kappa$

Hence $\lambda^2 = 1$. $\lambda = \pm 1$. Substituting into 6.5;

if $\lambda = -1$, then $1 + \kappa^2 = -2\kappa$ so $\kappa = -1$;

if $\lambda = 1$, then $1 + \kappa^2 = 2\kappa$ so $\kappa = 1$

**Definition:** Bosons are particles for which $\kappa = 1$, so that $\forall |x\rangle \in \mathbb{H}_0$ the creation operators $|x\rangle$ obey

6.6 $\quad\quad \forall y \in N \quad\quad |x;y\rangle = \frac{1}{\sqrt{2}}[(|x\rangle,|y\rangle) + (|y\rangle,|x\rangle)] = |y;x\rangle$

**Definition:** Fermions are particles for which $\kappa = -1$, so that $\forall |x\rangle \in \mathbb{H}_0$ the creation operators obey

6.7 $\quad\quad \forall y \in N \quad\quad |x;y\rangle = \frac{1}{\sqrt{2}}[(|x\rangle,|y\rangle) - (|y\rangle,|x\rangle)] = -|y;x\rangle$

The use of the ket notation for creation operators is justified by the homomorphism defined by

6.8 $\quad\quad |x\rangle|\rangle = \frac{1}{\sqrt{2}}[(|x\rangle,|\rangle) + \kappa(|\rangle,|x\rangle)]$

It is straightforward to check that this is a homomorphism with the scalar product defined by 5.5. In general the creation operator is defined by linearity

6.9 $\quad\quad \forall |f\rangle \in \mathbb{H} \quad\quad |f\rangle:\mathbb{H}^1 \to \mathbb{H}^2, \ |f\rangle = \sum_{x \in N} \langle x|f\rangle|x\rangle$

It follows immediately that $\forall |f\rangle, |g\rangle \in \mathbb{H}$

$|f;g\rangle = |f\rangle|g\rangle$

$= \sum_{x \in N} \langle x|f\rangle|x\rangle \sum_{y \in N} \langle y|g\rangle|y\rangle$

6.10 $\quad\quad = \sum_{x,y \in N} \langle x|f\rangle\langle y|g\rangle|x;y\rangle$

Using 6.10 gives

6.11 $\quad\quad \forall \text{ Boson } |f\rangle, |g\rangle \in \mathbb{H}^1 \quad\quad |f;g\rangle = |g;f\rangle$

and

6.12 $\quad\quad \forall \text{ Fermion } |f\rangle, |g\rangle \in \mathbb{H}^1 \quad\quad |f;g\rangle = -|g;f\rangle$

**Theorem:** The Pauli exclusion principle holds for fermions.

**Proof:** From 6.12, $\forall$ Fermion $|f\rangle \in \mathbb{H}^1 \quad |f;f\rangle = -|f;f\rangle$. Hence

6.13 $\quad\quad \forall \text{ Fermion } |f\rangle \in \mathbb{H}^1 \quad\quad |f;f\rangle = 0$

i.e. no two fermions may be in the same state.

The definition of the creation operator extends to $|x\rangle:\mathbb{H}^n \to \mathbb{H}^{n+1}$ by requiring that its action on each particle of an $n$ particle state is identical, and that it reduces to 6.1 in the restriction of $\mathbb{H}^n$ to the space of the $i$th particle. Thus $\forall x, y^i \in N, i = 1, ..., n$

6.14 $\quad\quad |x\rangle:(|y^1\rangle, ..., |y^n\rangle) \to \frac{1}{\sqrt{n+1}}\Bigg((|x\rangle,|y^1\rangle, ..., |y^n\rangle)$

$+ \kappa \sum_{i=1}^{n} (|y^i\rangle,|y^1\rangle, ..., |x\rangle, ..., |y^n\rangle)\Bigg)$

where $\kappa = 1$ for bosons and $\kappa = -1$ for fermions, and $|x\rangle$ appears in the $i+1$th position in the $i$th term of the sum. The normalisation is determined from 5.6 by observing that when all $x, y^i$ are distinct, the right hand side is the sum of $n+1$ orthogonal vectors, normalised to $\chi^{3(n+1)}$. 6.14 holds $\forall |f\rangle \in \mathbb{H}$ and $\forall |g\rangle \in \mathbb{H}^n$ by linearity.

**Definition:** The space of physically realisable states is the subspace $\mathscr{F} \subset \mathscr{H}$ which is generated from $\mathbb{H}^0 = \{|\,\rangle\}$ by the action of creation operators.

**Definition:** Notation for the elements of $\mathscr{F}$ is defined inductively.

$$6.15 \quad \forall \,|g\rangle \in \mathbb{H}^1, \forall \,|f\rangle \in \mathbb{H}^n \cap \mathscr{F} \quad |g\,;f\rangle = |g\rangle|f\rangle \in \mathbb{H}^{n+1} \cap \mathscr{F}$$

**Corollary:** $|g\,;f\rangle$ is identified with the creation operator $\mathscr{H} \to \mathscr{H}$ given by $|g\,;f\rangle = |g\rangle|f\rangle$

**Definition:** The bra corresponding to $|g\,;f\rangle \in \mathbb{H}^{n+1}$ is $\langle g\,;f|$.

**Theorem:** $\forall |x^i\rangle \in \mathbb{H}^1_0, i = 1, \ldots, n$

$$6.16 \quad |x^1\,;x^2\,;\ldots;x^n\rangle = \frac{1}{\sqrt{n!}} \sum_\pi \varepsilon(\pi)(|x^{\pi(1)}\rangle, \ldots, |x^{\pi(n)}\rangle)$$

where the sum runs over all permutations $\pi$ of $(1,2,\ldots,n)$, and $\varepsilon(\pi)$ is the sign of $\pi$ for fermions and $\varepsilon(\pi)=1$ for bosons.

**Proof:** By induction, 6.16 holds for $n = 2$, by 6.6 and 6.7. Now suppose that 6.16 holds $\forall n < m \in \mathbb{N}$, then, from definition 6.15

$$|x^1\,;x^2\,;\ldots;x^m\rangle = |x^1\rangle|x^2\,;\ldots;x^m\rangle$$
$$= \frac{1}{\sqrt{(m-1)!}} \sum_\pi \varepsilon(\pi)|x^1\rangle(|x^{\pi(2)}\rangle, \ldots, |x^{\pi(m)}\rangle)$$

by the inductive hypothesis. 6.16 follows from application of 6.14.

**Corollary:** $\forall \,|g\rangle, |f\rangle \in \mathbb{H}^1_0$ the creation operators obey the (anti)commutation relations

$$6.17 \quad [|g\rangle, |f\rangle]_\pm = 0$$

where for fermions

$$6.18 \quad [x, y]_+ = \{x, y\} = xy + yx$$

and for bosons

$$6.19 \quad [x, y]_- = [x, y] = xy - yx$$

**Proof:** By definition 6.15, $\forall x^i \in \mathbb{N}, i = 1, \ldots, n$, $\forall |x\rangle, |y\rangle \in \mathbb{H}^1_0$

$$|x\rangle, |y\rangle|x^1\,;x^2\,;\ldots;x^n\rangle = |x\,;y\,;x^1\,;x^2\,;\ldots;x^n\rangle$$
$$= \kappa|y\,;x\,;x^1\,;x^2\,;\ldots;x^n\rangle \quad \text{by 6.16}$$
$$= \kappa|x\rangle, |y\rangle|x^1\,;x^2\,;\ldots;x^n\rangle$$

But by definition the kets $|x^1\,;x^2\,;\ldots;x^n\rangle$ span $\mathscr{F}$. So by linearity

$$6.20 \quad [|x\rangle, |y\rangle]_\pm = 0$$

6.17 follows from 6.10.

**Corollary:** $\forall \,|f\rangle \in \mathbb{H}$ the creation operator $|f\rangle$ (anti)commutes with the creation operator $|\,\rangle$ for the vacuum state

$$6.21 \quad [|\,\rangle, |f\rangle]_\pm = 0$$

**Theorem:** $\forall |x^i\rangle, |y^i\rangle \in \mathbb{H}^1_0, i = 1, \ldots, n$

$$6.22 \quad \langle y^1\,;\ldots;y^n | x^1\,;\ldots;x^n\rangle = \sum_\pi \varepsilon(\pi) \prod_{i=1}^n \langle y^i | x^{\pi(i)}\rangle$$

**Proof:** By 6.16 and 5.5

$$\langle y^1;\ldots;y^n \mid x^1;\ldots;x^n \rangle = \frac{1}{n!}\sum_{\pi'}\varepsilon(\pi')\sum_{\pi''}\varepsilon(\pi'')\prod_{i=1}^{n}\langle y^{\pi'(i)} \mid x^{\pi''(i)}\rangle$$

$$= \frac{1}{n!}\sum_{\pi'}\varepsilon(\pi')\sum_{\pi\pi'}\varepsilon(\pi\pi')\prod_{i=1}^{n}\langle y^{\pi'(i)} \mid x^{\pi\pi'(i)}\rangle$$

where we observe that $\forall$ permutations $\pi''$, $\pi'$, $\exists$ a permutation $\pi$ such that $\pi'' = \pi\pi'$. 6.22 follows since the sum over $\pi'$ contains $n!$ terms which are identical up to the ordering of the factors in the product.

**Corollary:** $\forall \mid g_i\rangle, \mid f_j\rangle \in \mathbb{H}, i, j = 1, \ldots, n$

6.23 $$\langle g_1;\ldots;g_n \mid f_1;\ldots;f_n \rangle = \sum_{\pi}\varepsilon(\pi)\prod_{i=1}^{n}\langle g_i \mid f_{\pi(i)}\rangle$$

**Proof:** By linearity, 6.9, and definition 6.15

**Theorem:** $\forall n \in \mathbb{N}$, such that $0 < n$, $(\mathscr{F} \cap \mathbb{H}^n) \subset (\mathscr{F} \cap \mathbb{H}^{n+1})$ is an isomorphic embedding under the mapping $\mathbb{H}^n \to \mathbb{H}^{n+1}$ given by

6.24 $\quad\forall x^i \in \mathrm{N}, i = 1, \ldots, n \quad \mid x^1;\ldots;x^n\rangle \to \mid\rangle\mid x^1;\ldots;x^n\rangle = \mid;x^1;\ldots;x^n\rangle$

**Proof:** By 6.22 and 5.2

$$\langle ;y^1;\ldots;y^n \mid ;x^1;\ldots;x^n \rangle = \sum_{\pi \neq 1}\varepsilon(\pi)\langle\mid\rangle\prod_{i=2}^{n+1}\langle y^i \mid x^{\pi(i)}\rangle$$

$$= \langle y^1;\ldots;y^n \mid x^1;\ldots;x^n \rangle$$

by 5.1, and using 6.22 again.

## 7  Annihilation Operators

In an interaction particles may be created, as described by creation operators, and particles may change state or be destroyed. The destruction of a particle in an interaction is described by the action of an annihilation operator. A change of state of a particle can be described as the annihilation of one state and the creation of another, so a complete description of any process in interaction can be achieved through combinations of creation and annihilation operators. Annihilation operators incorporate the idea that it is impossible to tell which particle of a given type has been destroyed in the interaction. They are defined by their action on a basis of $\mathscr{H}$, and their relationship to creation operators will be determined. The use of bras to denote annihilation operators is justified by the obvious homomorphism defined below in 7.2 with $n = 1$.

**Definition:** $\forall \mid x\rangle \in \mathbb{H}_0^1$ the annihilation operator $\langle x \mid : \mathbb{H}^n \to \mathbb{H}^{n-1}$ $\langle x \mid : \mid f \rangle \to \langle x \mid f \rangle \in \mathbb{H}^{n-1}$ is given by $\forall x^i \in \mathrm{N}, i = 1, \ldots, n$

7.1 $\quad\langle x \mid\mid\rangle = \langle x \mid\rangle$

7.2 $\quad\langle x \mid (\mid x^1\rangle, \ldots, \mid x^n\rangle) = \frac{1}{\sqrt{n}}\sum_{i=1}^{n}\kappa^i\langle x \mid x^i\rangle(\mid x^1\rangle, \ldots, \mid x^{i-1}\rangle, \mid x^{i+1}\rangle, \ldots, \mid x^n\rangle)$

The normalisation in 7.2 is determined by observing that when all $x$, $x^i$ are distinct, the right hand side is the sum of $n$ orthogonal vectors, normalised to $\chi^n$ by 5.6. $\kappa = 1$ for bosons and $\kappa = -1$ for fermions, and is determined by considering the result of the annihilation operator on a one particle ket in $\mathbb{H}^1 \subset \mathbb{H}^n \cap \mathscr{F}$, which is identical for all values of $n$ under the isomorphic embedding of 6.24. The annihilation operator for any ket is defined by linearity

7.3 $\qquad \forall\, |f\rangle \in \mathbb{H} \quad \langle f|: \mathscr{F} \to \mathscr{F}$ is given by $\langle f| = \sum_{x \in N} \langle f|x\rangle\langle x|$

**Lemma:** $\forall |x\rangle, |x^1\rangle, |x^2\rangle \in \mathbb{H}_0^1$

7.4 $\qquad \langle x|(|x^1\rangle, |x^2\rangle) = \frac{1}{\sqrt{2}}\langle x|x^1\rangle|x^2\rangle + \kappa\langle x|x^2\rangle|x^1\rangle$

**Proof:** This is 7.2 with $n = 2$

**Theorem:** $\forall |y\rangle, |x^i\rangle \in \mathbb{H}_0^1, i = 1, \dots, n$

7.5 $\qquad \langle y|x^1;\dots;x^n\rangle = \sum_{i=1}^{n} \kappa^i \langle y|x^i\rangle |x^1;\dots;x^{i-1};x^{i+1};\dots;x^n\rangle$

**Proof:** By 6.16

$$\langle y\,||\,x^1;\dots;x^n\rangle = \langle y|\frac{1}{\sqrt{n!}}\sum_\pi \varepsilon(\pi)(|x^{\pi(1)}\rangle, \dots, |x^{\pi(n)}\rangle)$$

$$= \frac{1}{\sqrt{n}}\frac{1}{\sqrt{n!}}\sum_{i=1}^{n}\kappa^i\langle y|x^i\rangle n \sum_{\pi \neq i}\varepsilon(\pi)(|x^{\pi(1)}\rangle, \dots, |x^{\pi(n)}\rangle)$$

by 7.2, since for each value of $i \in \{1, \dots, n\}$ there are $n$ permutations $\pi$ which are identical apart from the position of $i$. 7.5 follows by applying 6.16 again.

**Theorem:** $\forall |x^i\rangle, |y^i\rangle \in \mathbb{H}_0^1, i = 1, \dots, n$

7.6 $\qquad \langle y^n|\dots\langle y^1||x^1;\dots;x^n\rangle = \langle y^1;\dots;y^n | x^1;\dots;x^n\rangle$

**Proof:** From 6.1

$$\langle y^1||x^1;x^2\rangle = \tfrac{1}{2}\sqrt{2}\langle y^1|[(|x^1\rangle, |x^2\rangle) + \kappa(|x^2\rangle, |x^1\rangle)]$$

$$= \tfrac{1}{2}((1+\kappa^2)\langle y^1|x^1\rangle|x^2\rangle + 2\kappa\langle y^1|x^2\rangle|x^1\rangle)$$

by applying 7.4. Then

$$\langle y^2|\langle y^1||x^1;x^2\rangle = \tfrac{1}{2}((1+\kappa^2)\langle y^1|x^1\rangle\langle y^2|x^2\rangle + 2\kappa\langle y^1|x^2\rangle\langle y^2|x^1\rangle)$$

$$= \langle y^1|x^1\rangle\langle y^2|x^2\rangle + \kappa\langle y^1|x^2\rangle\langle y^2|x^1\rangle$$

$$= \langle y^1;y^2|x^1;x^2\rangle$$

by 6.22. So 7.6 holds for $n = 2$.

Now suppose 7.6 holds for $n < m \in \mathbb{N}$ and apply $\langle y^m|\dots\langle y^2|$ to 7.5

$$\langle y^m|\dots\langle y^2|\langle y^1|x^1;\dots;x^m\rangle = \sum_{i=1}^{m}\kappa^i\langle y^1|x^i\rangle\langle y^2;\dots;y^m|x^1;\dots;x^{i-1};x^{i+1};\dots;x^m\rangle$$

$$= \sum_{i=1}^{m}\kappa^i\langle y^1|x^i\rangle\sum_{\pi \neq i}\varepsilon(\pi)\prod_{i=2}^{m}\langle y^i|x^{\pi(i)}\rangle \qquad \text{by 6.22}$$

$$= \sum_\pi \varepsilon(\pi)\prod_{i=1}^{m}\langle y^i|x^{\pi(i)}\rangle \qquad \text{since all } m \text{ terms are identical}$$

$$= \langle y^1;\dots;y^m | x^1;\dots;x^m\rangle$$

by 6.22. So 7.6 holds for $\forall n \in \mathbb{N}$ by induction.

**Corollary:** $\forall |x^i\rangle \in \mathbb{H}_0^1, i = 1, ..., n \;\; \forall |f\rangle \in \mathbb{H}^n \cap \mathscr{F}$

$$\langle x^1;...;x^n|f\rangle = \langle x^n|...\langle x^1||f\rangle$$

**Proof:** Immediate from 7.6, by linearity. Hence, it is consistent to define:

**Definition:** $\forall |x^i\rangle \in \mathbb{H}_0^1, i = 1, ..., n$ the annihilation operator $\langle x^1;...;x^n|: \mathscr{H} \to \mathscr{H}$ is given by

7.7 $\qquad \langle x^1;...;x^n| = \langle x^n|...\langle x^1|$

**Definition:** On a complex vector space, $\mathscr{V}$, with a hermitian form, the hermitian conjugate $\phi^\dagger: \mathscr{V} \to \mathscr{V}$ of the linear operator $\phi: \mathscr{V} \to \mathscr{V}$ is defined such that $\forall f, g \in \mathscr{V}. (\phi^\dagger f, g) = (f, \phi g)$. It is routine to show that $\phi^\dagger$ is a linear operator.

**Theorem:** $\forall |x^i\rangle \in \mathbb{H}_0^1, i = 1, ..., n$ the creation operator $|x^1;...;x^n\rangle: \mathscr{F} \to \mathscr{F}$ is the hermitian conjugate of the annihilation operator, $\langle x^1;...;x^n|: \mathscr{F} \to \mathscr{F}$.

7.8 $\qquad \langle x^1;...;x^n| = |x^1;...;x^n\rangle^\dagger$

**Proof:** From the definition, $\forall x^i, y^j \in N, i = 1, ..., n, j = 1, ..., m \;\; \forall |f\rangle \in \mathscr{F},$

$$\langle y^1;...;y^n|\langle x^1;...;x^n|^\dagger|f\rangle = \langle y^1;...;y^n|\langle x^1;...;x^n||f\rangle$$
$$= \langle y^n|...\langle y^1|\langle x^n|...\langle x^1||f\rangle$$
$$= \langle x^1;...;x^n;y^1;...;y^n||f\rangle$$

by applying 7.7 three times. Thus $\langle x^1;...;x^n|^\dagger$ is the map

$$\langle x^1;...;x^n|^\dagger: |y^1;...;y^n\rangle \to |x^1;...;x^n;y^1;...;y^n\rangle$$

which demonstrates 7.8.

**Corollary:** $\forall |g\rangle, |f\rangle \in \mathbb{H}$ the annihilation operators obey the (anti)commutation relations.

7.9 $\qquad [\langle g|, \langle f|]_\pm = 0$

**Proof:** Straightforward from, 6.17, the (anti)commutation relations for creation operators.

**Theorem:** $\forall |g\rangle, |f\rangle \in \mathbb{H}$ the creation operators and annihilation operators obey the (anti)commutation relations

7.10 $\qquad [\langle g|, |f\rangle]_\pm = \langle g|f\rangle$

**Proof:** By 7.5, $\forall |y\rangle, |x^i\rangle \in \mathbb{H}_0^1, i = 1, ..., n$

$$\langle y|x;x^1;...;x^n\rangle = \sum_{i=1}^n \kappa^{i+1}\langle y|x^i\rangle |x;x^1;...;x^{i-1};x^{i+1};...;x^n\rangle$$
$$+ \langle y|x\rangle |x^1;...;x^n\rangle$$

Using 7.5 again

$$\langle y||x\rangle |x^1;...;x^n\rangle = \kappa|x\rangle\langle y|x^1;...;x^n\rangle + \langle y|x\rangle |x^1;...;x^n\rangle$$

Therefore

7.11 $\qquad [\langle y|, |x\rangle]_\pm = \langle y|x\rangle$

7.10 follows from 6.9 and 7.3.

**Corollary:** $\forall |f\rangle \in \mathbb{H}$ the annihilation operator obeys the (anti)commutation relation

7.12 $\qquad [|\rangle, \langle f|]_\pm = 0$

## 8   Classical Correspondence

Real measurements do not achieve an accuracy in the order of chronons. In a measurement of position, the ket, $|f\rangle$, naming the initial state of the apparatus is changed into a ket named by a position in $X$, where $X$ is a region of space determined by the measuring apparatus. The operator effecting the change is

8.1 $$Z(X) = \sum_{x \in X} |x\rangle\langle x|$$

as is shown by direct application

8.2 $$Z(X)|f\rangle = \sum_{x \in X} |x\rangle\langle x|f\rangle$$

since the resulting ket is a weighted logical or between positions in $X$. Applying $Z$ a second time gives

$$Z(X)Z(X)|f\rangle = \sum_{y \in X} |y\rangle\langle y| \sum_{x \in X} |x\rangle\langle x|f\rangle$$

$$= \sum_{y \in X} |y\rangle\langle y|f\rangle \qquad \text{by 3.1}$$

So $Z(X)$, is a projection operator

8.3 $$Z(X)Z(X) = Z(X) \qquad \text{by 8.1}$$

reflecting the observation that a second measurement of a quantity gives the same result as the first (see e.g. [14]). By applying 8.3 to the ket $|f\rangle$ normalised so that $\langle f|f\rangle = 1$ obtain

8.4 $$\langle f|Z(X)Z(X)|f\rangle = \sum_{x \in X} \langle f|x\rangle\langle x|f\rangle$$

But 8.4 is the sum of the probabilities that the particle is found at each individual position, $x \in X$. In other words it is the probability that a measurement of position finds the particle in the region $X$. In the case that $X$ contains only the point $x$, $X = \{x\}$, 8.2 becomes

8.5 $$Z(x)|f\rangle = |x\rangle\langle x|f\rangle$$

Thus, the position function, $\langle x|f\rangle$, can be reinterpreted as the magnitude of the projection, $Z(x)$, from the ket $|f\rangle$ of the apparatus into the ket $|x\rangle$.

$Z(X)$, is not simply a mathematical device to produce a result; it summarises the physical processes taking place in the interactions involved in a measurement of position. If a measurement of position performed on $|f\rangle$ has resulted in a position in $X$, $Z(X)$ has, in effect, been applied to $|f\rangle$ and it is assumed that $Z(X)$ is generated by some combination of particle interactions. Classical probability theory describes situations in which every parameter exists, but some are not known. Probabilistic results come from different values taken by unknown parameters. We have a similar situation here. There are no relationships between particles apart from those generated by interactions. The configuration of particles has been largely determined by setting up the experimental apparatus, but the precise pattern of interactions is unknown. Thus the current model describes a classical probability in which the unknowns lie in the configuration of interacting particles. The probability that the interactions combine to $Z(x)$ is given by

8.6 $$\langle f|Z(x)Z(x)|f\rangle = \langle f|Z(x)|f\rangle = \langle f|x\rangle\langle x|f\rangle = |\langle x|f\rangle|^2$$

Thus, 3.3 can be understood as a classical probability function, where the variable, $x$, runs over the set of projection operators,

8.7 $$Z(x) = |x\rangle\langle x|$$

such that each $Z(x)$ is generated by an unknown configuration of particle interactions in measurement.

In general, measurements generate numerical values and are repeated many times over from the same starting state. Then the average value of the result is taken. Expectation is the term used in statistics for the prediction of an average value. Under the laws of statistics, the more repetitions, the closer the average value will be to the expectation of the measurement. By the definition of expectation in statistics, if $O(x)$ is a real function of a position, $x \in \mathrm{N}$, then, given the ket $|f\rangle \in \mathbb{H}$, $x$ is a random variable with probability function $|\langle f|x\rangle|^2$. and the expectation of $O(x)$ is

8.8 $$\langle O \rangle = \sum_{x \in X} \langle f|x\rangle O(x) \langle x|f\rangle$$

It is straightforward to generalise the analysis to let $O(x)$ be a real valued functional. If we define an operator on $\mathscr{F}$ by the formula

8.9 $$O = \sum_{x \in X} |x\rangle O(x) \langle x|$$

then the expectation of $O$ given the initial ket $|f\rangle \in \mathbb{H}$ is

8.10 $$\langle O \rangle = \langle f|O|f\rangle \qquad \text{by 8.8 and 8.9}$$

By 7.5 $O$ is additive for independent multiparticle states, so 8.8 applies also to the expectation for all $|f\rangle \in \mathscr{F}$. $O$ is hermitian, so there is a particular class of kets, called eigenkets, such that if $|f\rangle$ is an eigenket, then $\exists r \in \mathbb{R}$ known as the eigenvalue associated with $|f\rangle$ such that

8.11 $$O|f\rangle = r|f\rangle.$$

If all physical processes are described by a composition of interaction operators then the existence of an observable quantity depends not on whether an observation takes place, but on the configuration of matter. If the interaction operators describing a physical process combine to generate a hermitian operator, then the corresponding observable quantity exists, independent of observation or measurement. The state is said to be an eigenstate of the observable, and is labelled by an eigenket of the operator. The value of the observable quantity is given by the corresponding eigenvalue

8.12 $$\langle O \rangle = \langle f|O|f\rangle = \langle f|r|f\rangle = r\langle f|f\rangle = r$$

We know from experiment that measurements generate definite results, and thereby provide definite categorisations of states by means of a kets. This is equivalent to the application of a projection operator. In a statistical analysis of a large number of particles, each result names a physical process described by a combination of operators equivalent to a projection operator. Under the identification of addition with logical OR the expectation of all the results is a hermitian operator equal to a weighted sum over a family of projection operators. Classical laws are derived from the expectation, 8.12, of the interactions of large numbers of particles.

## 9    Discrete Wave Mechanics

Let $\mathrm{T} \subset \mathbb{N}$ be a finite discrete time interval such that any particle under study certainly remains in N for $x_0 \in \mathrm{T}$. Without loss of generality let $\mathrm{T} = [0, T)$.

**Definition:** An interaction at time $t$ is described by an operator, $I(t): \mathscr{F} \to \mathscr{F}$. For definiteness we may take

9.1 $$\forall x^i \in \mathrm{N}, \forall n \in \mathbb{N}, \langle x^1;\ldots;x^n|I|x^1;\ldots;x^n\rangle = 0$$

since otherwise there would be a component corresponding to the absence of interaction.

At each time $t$, either no interaction takes place and the state $|f\rangle \in \mathscr{F}$ is unchanged, or an interaction, $I$, takes place. By the identification of the operations of vector space with weighted OR between uncertain possibilities, the possibility of an interaction at time $t$ is described by the map $\mathscr{F} \to \mathscr{F}$

$$|f\rangle \to \mu(1 - iI(t))|f\rangle$$

where $\mu$ is a scalar value chosen to preserve the norm, as required by the probability interpretation. Thus the law of evolution of the ket in one chronon from time $t$ to time $t + 1$ is

9.2 $\qquad |f\rangle_{t+1} = \mu(1 - iI(t))|f\rangle_t$

In the absence of interaction, the ket for a particle at rest does not change (such states exist as a particle is always at rest in its own reference frame). So $\forall |f\rangle \in \mathbb{H} \ \exists \lambda \in \mathbb{C}$ such that 9.2 reduces to

9.3 $\qquad |f\rangle_{t+1} = \lambda |f\rangle_t$

Preservation of the norm implies that $\exists m \in \mathbb{R}$ such that $\lambda = e^{im}$, so that

9.4 $\qquad |f\rangle_{t+1} = e^{im}|f\rangle_t$

**Definition:** $m$ is the bare mass of a particle. It will be found that $m$ can be identified with the classical concept of mass.

Then 9.3 is a geometric progression with solution

9.5 $\qquad |f\rangle_t = e^{imt}|f\rangle_0$

and by 3.1, the solution for a particle in its own reference frame is

9.6 $\qquad \langle t, \mathbf{x} | \mathbf{0} \rangle = e^{imt}\delta_{\mathbf{x}\mathbf{0}}$

A particle's reference frame is not directly observable, but we can use quantum transformation to find the solution in a macroscopic frame.

**Definition:** For any single particle ket, $|f\rangle$, normalised so that $\langle f|f\rangle = 1$ the quantum transformation $Q(f)$ is given by

9.7 $\qquad Q(f) = \int_M d^3p\, |\mathbf{p}\rangle\langle \mathbf{p}|f\rangle\langle \mathbf{p}|$

**Theorem:**

9.8 $\qquad Q(f)|\mathbf{0}\rangle = |f\rangle$

**Corollary:** $Q(f)$ transforms the particle's reference frame to a macroscopic frame.
**Proof:**

9.9 $\qquad Q(f)|\mathbf{x}\rangle = \int_M d^3p\, |\mathbf{p}\rangle\langle \mathbf{p}|f\rangle\langle \mathbf{p}|\mathbf{x}\rangle = \left(\frac{1}{2\pi}\right)^{\frac{3}{2}} \int_M d^3p\, |\mathbf{p}\rangle e^{i\mathbf{x}\cdot\mathbf{p}}\langle \mathbf{p}|f\rangle$

9.8 follows by letting $\mathbf{x} = \mathbf{0}$.

**Corollary:** $Q$ is not unitary.
**Proof:**

9.10 $\qquad Q^\dagger(f)|f\rangle = \int_M d^3p\, |\mathbf{p}\rangle\langle f|\mathbf{p}\rangle\langle \mathbf{p}|f\rangle \ne |\mathbf{0}\rangle$

Thus information is lost in a quantum transform, reflecting the lost information in a macroscopic frame.

**Corollary:** For a clock with a certain position $z \in \mathrm{N}$, $Q(z)$ is a space translation
**Proof:**

$$9.11 \qquad Q(z) = \int_M d^3p |p\rangle\langle p|z\rangle\langle p| = \left(\frac{1}{2\pi}\right)^{\frac{3}{2}} \int_M d^3p |p\rangle e^{iz \cdot p} \langle p|$$

since $\langle x|Q(z)|f\rangle = \langle x-z|f\rangle$ by 4.13.

## 10 Continuous Wave Mechanics

From 4.6, at any time $t \in \mathrm{T}$

$$10.1 \qquad \langle x|f\rangle = \left(\frac{1}{2\pi}\right)^{\frac{1}{2}} \int_M dp \, \langle p|f\rangle \, e^{-ix \cdot p}$$

Although $\langle x|f\rangle$ is, by definition, discrete, on a macroscopic time-scale it appears continuous, and it is natural to embed $\langle x|f\rangle$ into a continuous function $f(x)$, by simply replacing $x$ with a continuous variable

$$10.2 \qquad f(x) = \left(\frac{1}{2\pi}\right)^{\frac{1}{2}} \int_M dp \, \langle p|f\rangle \, e^{-ix \cdot p}$$

Similarly 9.6 can be embedded into a continuous function of time $f:\mathbb{R} \to \mathbb{C}$ given by

$$10.3 \qquad f(t) = e^{imt}\delta_{x0}$$

It is worth observing at this point that this discrete reformulation of quantum mechanics is fully consistent with general relativity, since the braket is defined in a finite region. Let $\Omega$ be a differentiable manifold with $n$ dimensions. Let $g$ be the metric tensor on $\Omega$. We have $x \cdot p = x_i g^{ij} p_j$, where I have used the standard convention on raising and lowering indices. We seek a continuous function, $f:\Omega \to \mathbb{C}$, called the wave function, such that, if the particle certainly remains in N for time interval, $\mathrm{T} \subset \mathbb{N}$, then $\langle x|f\rangle$ can be embedded into $f(x)$

$$10.4 \qquad \forall x \in \mathrm{N}, \forall t \in \mathrm{T} \quad \langle x|f\rangle = \langle t, x|f\rangle = f(t,x) = f(x)$$

Physical law will be expressed in terms of creation and annihilation operators, and must also be Lorentz covariant. Lorentz transformation cannot be applied directly to functions of a discrete co-ordinate system. But it must be applied to the creation and annihilation operators appearing in the interaction operator, and hence to the wave function by homomorphism. Then 10.4 defines a position function, and hence a ket, by the restriction of the wave function to N in the transformed co-ordinate system at integer time. For any ket, there is a unique momentum space function defined by 4.4, and a unique wave function defined by 10.2. So there is a homomorphism between $\mathbb{H}$ and the vector space of wave functions with the hermitian product defined by 3.5. Wave functions are not restricted to $\mathscr{L}^2$, and 3.5 is not the hermitian product of Hilbert space, but by the definition of convergence of an integral, it is approximated by the hermitian product whenever $f$ and $g$ are in $\mathscr{L}^2$ and $\chi$ can be regarded as small.

Invariance under Lorentz transformation requires that the law of time evolution has a Lorentz invariant form when expressed in terms of wave functions. The law for the time evolution of the wave function for a stationary particle is given by differentiating 10.3 with respect to time

$$10.5 \qquad -i\partial_0 f = mf$$

Then 9.4 is obtained by integrating 10.5 over one chronon. Thus, in the restriction to integer values, 10.5 is identical to 9.4, the difference equation for a stationary non-interacting particle. It is therefore an expression of the same relationship or law. As an equation of the wave function, the right hand side of 10.5 is a scalar, whereas the left hand side is the time component of a vector whose space component is zero. So 10.5 is not manifestly covariant. For a covariant equation which reduces to 10.5 for a particle in its own reference frame, we take a scalar product involving the vector derivative, $\partial$, and the wave function

10.6 $\qquad -i\partial \cdot \Gamma f = mf$

Then the time evolution of the position function in any reference frame is the restriction of the solution of 10.6 to N at time $t \in \mathrm{T}$. As discovered by Dirac [15], there is no invariant equation in the form of 10.6 for scalar $f$ and the theory breaks down. The simplest known solution requires that we redefine N to have 3 dimensions plus spin indexed by the set $S$

10.7 $\qquad \mathrm{N} = (-\nu, \nu] \otimes (-\nu, \nu] \otimes (-\nu, \nu] \otimes S \subset \mathbb{N}^3 \otimes S$ for some $\nu \in \mathbb{N}$.

Now the constructions of the vector spaces, $\mathbb{H}$, $\mathscr{H}$ and $\mathscr{F}$, go through as before, and when we wish to make the spin index explicit we write

10.8 $\qquad |x\rangle = |x, \alpha\rangle = |x\rangle_\alpha$

normalised by 3.1 so that

10.9 $\qquad \forall (x, \alpha), (y, \beta) \in \mathrm{N}_S \quad \langle x, \alpha | y, \beta \rangle = \langle x | y \rangle_{\alpha\beta} = \delta_{xy}\delta_{\alpha\beta}$

The wave function acquires a spin index

10.10 $\qquad f(x) = f_\alpha(x) = \langle x | f \rangle_\alpha$

and the braket becomes

10.11 $\qquad \langle g | f \rangle = \sum_{x \in \mathrm{N}} \langle g | x \rangle \langle x | f \rangle = \sum_{(\mathbf{x}, \alpha) \in \mathrm{N}} \overline{g_\alpha(x)} f_\alpha(x)$

It is now possible to find a covariant equation which reduces to 10.5 in the particle's reference frame, namely the Dirac equation,

10.12 $\qquad i\partial \cdot \gamma f(x) = mf(x)$

Another possibility is that $f$ is a vector and that 10.5 is a representation of a vector equation with $m = 0$

10.13 $\qquad i\partial \cdot f(x) = 0$

As is well known the requirement for Lorentz covariance of 10.6 is related to the homogeneity of the universe. It applies in the current model, as it does in the standard model and now suggests an answer to the question "why are there three dimensions?". On its own the behaviour of particles placed no requirement on the number of dimensions. But when ket space is introduced as a means of analysing physical law it is found that the simplest solution requires three dimensions. It appears that four dimensional space-time is the simplest environment which can be built from the interactions of point-like particles.

# A Pre-Geometric Model Exhibiting Physical Law 20